\newif\ifARXIV
\ARXIVtrue

\documentclass[sigconf]{acmartarxiv}

\def\BibTeX{{\rm B\kern-.05em{\sc i\kern-.025em b}\kern-.08emT\kern-.1667em\lower.7ex\hbox{E}\kern-.125emX}}

\ifARXIV
        \setcopyright{none}
\fi

\newif\iflongversion

\longversionfalse

\usepackage[toc,page]{appendix}
\usepackage{nicefrac}
\usepackage{multirow}

\usepackage{amsfonts,latexsym}
\usepackage{amsthm}
\usepackage{amsmath}
\usepackage{amssymb}
\usepackage{caption}
\usepackage{verbatim}

\usepackage{resizegather}
\usepackage{appendix}
\usepackage{graphicx,pifont,color}

\usepackage[ruled,vlined]{algorithm2e}
\usepackage{algpseudocode}
\usepackage{hhline}
\usepackage{cleveref}
\usepackage{lhelp}

\usepackage{fancyvrb}
\usepackage{lipsum}
\usepackage{pstricks,pst-node,pst-grad,pst-plot}

\usepackage{multirow}

\usepackage{paralist}

\usepackage{float}
\usepackage{subcaption}
\usepackage{tikz}

\usepackage{afterpage}

\usepackage{pgfplots}
\usetikzlibrary{positioning,calc}
\usetikzlibrary{arrows.meta,chains,decorations.pathreplacing,scopes,positioning}
\usetikzlibrary{calc,shapes.multipart,chains,arrows}

\usepackage{lhelp}
\usepackage{listings}
\usepackage{float}
\usepackage{xcolor}
\usepackage{etoolbox}

\definecolor{commentgreen}{RGB}{2,112,10}
\definecolor{eminence}{RGB}{158,58,180}
\definecolor{weborange}{RGB}{230,145,0}
\definecolor{frenchplum}{rgb}{0.55, 0.0, 0.55}

\lstdefinestyle{mystyle} {
	language=C++,
	frame=tb,
	tabsize=4,
	showstringspaces=false,
	numbers=left,
	numbersep=1pt,
	backgroundcolor=\color{black!3}, 
	belowcaptionskip=\parskip,
	breaklines=true,
	basicstyle=\tiny\ttfamily,
	commentstyle=\color{commentgreen},
	keywordsprefix=\&,alsoletter=\&,%
	keywordstyle=\bfseries\color{eminence},
	numberstyle=\tiny\color{gray},
	stringstyle=\color{weborange},
	basicstyle=\scriptsize\ttfamily, 
	emph={int,char,double,float,unsigned,void,bool,AltArrZ\_t, mpz\_t, degrees\_t, SMZP, Poly, RegChain, AsyncGen, Poly, RC, Object},
	emphstyle={\color{blue!95!black}},
	literate={\&}{{{\&}}}1,
	otherkeywords={>,<,.,;,-,!,=,~},
	morekeywords = [2]{>,<,.,;,-,!,=,~},
	keywordstyle = [2]\color{black}
}

\usepackage{multicol}
\usepackage[utf8]{inputenc}

\makeatletter
\DeclareRobustCommand*\cal{\@fontswitch\relax\mathcal}
\makeatother

\def\K{\mbox{${\mathbb K}$}}
\def\L{\mbox{${\mathbb L}$}}

\def\Q{\mbox{${\mathbb Q}$}}

\def\Z{\mbox{${\mathbb Z}$}}

\def\0 {\ensuremath{\mathbf{0}}}

\newcommand{\maple}{{\sc Maple}}

\newcommand{\cilk}{\mbox{\sc Cilk}}
\newcommand{\cpp}{\mbox{\sc C++}}

\newcommand{\openmp}{\mbox{\sc OpenMP}}

\algnewcommand\algorithmicswitch{\textbf{switch}}
\algnewcommand\algorithmiccase{\textbf{case}}
\algnewcommand\algorithmicassert{\texttt{assert}}
\algnewcommand\Assert[1]{\State \algorithmicassert(#1)}%
\algdef{SE}[SWITCH]{Switch}{EndSwitch}[1]{\algorithmicswitch\ #1\ \algorithmicdo}{\algorithmicend\ \algorithmicswitch}%
\algdef{SE}[CASE]{Case}{EndCase}[1]{\algorithmiccase\ #1}{\algorithmicend\ \algorithmiccase}%
\algtext*{EndSwitch}%
\algtext*{EndCase}%

\SetKwProg{CilkFor}{cilk\_for}{ {}do}{}
\SetKwProg{ParFor}{parallel\_for}{ {}do}{}

\newcommand{\hidetext}[1]{\mbox{ \ }}

\newcommand{\textblockcomment}[1]{}

\newcommand{\Regularize}[1]{\mbox{{\sf Regularize}$(#1)$}}

\newcommand{\Intersect}[1]{\mbox{{\sf Intersect}$(#1)$}}
\newcommand{\Triangularize}[1]{\mbox{{\sf Triangularize}$(#1)$}}

\newcommand{\RemoveRedundantComponents}[1]{{{\sf Remove\-Redundant\-Components}(#1)}}
\newcommand{\TriangularizeLevel}[1]{{{\sf Triangularize\-Level}(#1)}}
\newcommand{\TriangularizeBubble}[1]{{{\sf Triangularize\-Bubble}(#1)}}

\newcommand{\SolveSystemIncrementally}[1]{\mbox{{\sf SolveSystemIncrementally}$(#1)$}}

\newcommand{\SolveOneEquation}[1]{\mbox{{\sf SolveOneEquation}$(#1)$}}

\begin{document}
\title{On the Parallelization of Triangular Decomposition of Polynomial Systems}

\author{Mohammadali Asadi}
\affiliation{\institution{University of Western Ontario, London, Canada}}
\email{masadi4@uwo.ca}

\author{Alexander Brandt}
\affiliation{\institution{University of Western Ontario, London, Canada}}
\email{abrandt5@uwo.ca}

\author{Robert H.C. Moir}
\affiliation{\institution{University of Western Ontario, London, Canada}}
\email{robert@moir.net}

\author{Marc~Moreno~Maza}
\affiliation{\institution{University of Western Ontario, London, Canada}}
\email{moreno@csd.uwo.ca}

\author{Yuzhen Xie}
\affiliation{\institution{University of Western Ontario, London, Canada}}
\email{yuzhenxie@yahoo.ca}

\renewcommand{\shortauthors}{Asadi, Brandt, Moir, Moreno~Maza, Xie}

\begin{abstract}
We discuss the parallelization of algorithms for solving polynomial
systems symbolically by way of triangular decomposition.  
Algorithms for solving polynomial systems combine low-level routines
for performing arithmetic operations on polynomials and high-level
procedures which produce the different components (points, curves,
surfaces) of the solution set.
The latter ``component-level'' parallelization of triangular
decompositions, our focus here,
belongs to the class of dynamic irregular parallel applications.
Possible speed-up factors depend on geometrical properties of the
solution set (number of components, their dimensions and degrees);
these algorithms do not scale with the number of processors.
\\
In this paper we combine two different concurrency schemes, the
fork-join model and producer-consumer patterns, to better capture
opportunities for component-level parallelization. We report on our
implementation with the publicly available BPAS library. Our
experimentation with 340 systems yields promising results.
\end{abstract}


%




\maketitle

\setlength\textfloatsep{\parskip}

\section{Introduction}

In linear algebra, stencil computations, sorting algorithms, fast
Fourier transform and other standard kernels in scientific computing,
opportunities for parallel computations often come from either a
divide-and-conquer scheme, with concurrent recursive calls, or a
for-loop nest where some iterations can be executed concurrently.

In the case of {\em non-linear} polynomial algebra, opportunities for
parallel computations can be categorized between low-level routines
for performing arithmetic operations on polynomials and high-level
procedures for algebraic and geometric computations.
Many of those low-level routines
follow patterns that are similar to standard kernels in scientific
computing and their parallelization has been well studied
for both dense and sparse polynomials~\cite{monagan2009parallel,monagan2010parallel,DBLP:conf/casc/GastineauL13,DBLP:conf/cap/GastineauL15}.

The parallelization of high-level procedures for algebraic and
geometric computations received much attention in the 80's and 90's,
see~\cite{buchberger1987parallelization,saunders1989parallel,bungden1994fine,faugere1994parallelization,attardi1996strategy}.
More recently in~\cite{maza2007component}, two of the current
authors consider 
the parallelization of
triangular decompositions of polynomial systems.

Triangular decomposition---a symbolic method for solving 
systems of polynomial equations---uses
algebraic techniques to split the solution
space into geometric components (i.e. points, curves, surfaces, etc.).
For those techniques,
opportunities for parallel computations depend only on the geometry of the
input system and may vary from none to many. Moreover, when such
opportunities are present the workload between tasks may be largely
unbalanced. 

\subsection{Problem Statement}

The preliminary work reported in~\cite{maza2007component} does
not cover all opportunities for concurrent/parallel processing of {\em
	component-level}, i.e. high-level, procedures in triangular
decompositions of polynomial systems, leaving open the full question of
possibilities for component-level parallelism,
both in terms of algorithm scheme and implementation techniques.  
The tests also used only a few examples
computed modulo a prime number instead of using
rational number coefficients, increasing the number of 
the components and hence the possible speed-up factors with respect to serial.


To stress the importance of the proposed question one can make the
following observation. In computer algebra systems like {\maple} only
low-level routines for polynomial arithmetic use multithreaded
parallelism 
while high-level procedures, such as {\maple}'s {\tt
	solve} command (for solving systems of systems of polynomial or
differential equations symbolically) execute serial code.

\subsection{Contributions}

We make use of both the fork-join model and the pipeline pattern (via asynchronous producer-consumer)
to exploit opportunities for parallelism in the component-level of triangular decomposition.  To
our knowledge, this is the first time that the producer-consumer model
is used in a high-level algorithm in symbol computation.

Triangular decompositions have two important features: the necessity
of removing {\em redundant} (that is, superfluous) components and the
possibility of avoiding the computation of {\em degenerate
	components}.  These features lead to different algorithm variants
that we call {\em Level} vs. {\em Bubble} (two different strategies
for removing redundant components), and {\em Lazard-Wu} vs. {\em
	Kalkbrener} decompositions (that is whether or not degenerate
components may be ignored).

These different algorithm variants combined with three degrees of parallelism
granularity (fine-grained, coarse-grained, and serial) lead to 12
different configurations of our implementation. 
The parallelism in our implementation is achieved using the standard C++11
thread support library as well as the Cilk extension of C/C++.

We test the various configurations of our implementation by
considering a suite of approximately 3000 real-world polynomial
systems.  These systems, provided by MapleSoft (the company developing
{\maple}), come from a combination of actual user data, bug reports,
and the scientific literature.  We have obtained speed-up factors up to
8 on a 12-core machine.  We highlight that the underlying polynomial
arithmetic is performed serially since we focus on component-level parallelization 
of triangular decomposition.
Among these configurations we find that the Bubble variant admits more parallelism than the Level
variant while solving in the Kalkbrener sense admits more parallelism than solving in the Lazard-Wu sense.
By this, we mean that Bubble and Kalkbrener benefit from parallelism
more often and to a higher a degree.

\subsection{Structure of the Paper}

Following a presentation of the mathematical background
for triangular decomposition in Section~\ref{sec:Background}, 
Section~\ref{sec:Triangularize} reviews the main
features of triangular decomposition computations.
Section~\ref{sec:practice} discusses opportunities for 
concurrent execution in these computations.
Section~\ref{sec:Implementation} reports on their implementation, 
and Section~\ref{sec:Experimentation}
presents our experimentation.
Section~\ref{sec:future} highlights directions for future work.

\section{Algebraic background}
\label{sec:Background}

Section~\ref{sec:TheD5Principle} focuses on a celebrated algebraic
construction by which computations can split when solving polynomial
systems, thus bringing opportunities for parallel computations, though
not necessarily balanced in terms of workload.  This construction,
called the {\em D5 Principle} after its authors~\cite{della1985new} relies on
the well-known Chinese remainder theorem, reviewed in
Section~\ref{sec:CRT}.  

\subsection{Solving polynomial systems incrementally}
\label{sec:solvingincrementally}

An informal sketch of the top-level procedure considered in this paper
is given by Algorithm~\ref{Algo:SolveIncrementally}, while a more formal
version is stated as Algorithm~\ref{Algo:Triangularize}.  For a finite
set $F$ of polynomial equations, the function
\SolveSystemIncrementally{F} returns the common solutions of the 
members of $F$ described as a finite set of {\em components}
$T_1, \ldots, T_e$.  By a component, we mean a set of polynomial
equations with remarkable algebraic properties; see
Section~\ref{sec:Triangularize} for more details.
Algorithm~\ref{Algo:SolveIncrementally} proceeds by incrementally solving
one equation after another against each component produced by the
previously solved equations.  The core routine is
$\SolveOneEquation{p, T}$ which solves the polynomial equation $p$
against the component $T$, that is, finding the solutions
of $p$ that are also solutions of $T$.

\begin{algorithm}
	\caption{\SolveSystemIncrementally{F}}
	\label{Algo:SolveIncrementally}
	\KwIn{a finite set $F$ of polynomial equations}
	\KwOut{a set of components such that their union is
		the solution set of $F$}
	{\bf if} $|F|==0$ {\bf then} \textbf{return} $[\{\}]$\;
	Choose a polynomial $p\in F$\;
	\For{$T {\bf\ in\ } \SolveSystemIncrementally{F\setminus\{p\}}$}{
		{\bf yield} $\SolveOneEquation{p, T};$
	}
\end{algorithm}

\begin{example}
	\label{ex:1}
	Consider the polynomial system:
	\begin{equation*}
	\left\{
	\begin{array}{rl}
	f_1: & x^3 - 3x^2 + 2x =  0 \\
	f_2: & 2yx^2 - x^2 -3yx + x  =  0 \\
	f_3: & zx^2 - zx =  0 
	\end{array} \right.
	\end{equation*}
	Calling $\SolveOneEquation{f_1, \varnothing}$
	yields $\{ f_1 \}$ since $f_1$ is ``simple enough''
	to define a component on its own, namely the
	points $(x, y, z)$ satisfying $x^3 - 3x^2 + 2x =  0$.
	
	Consider now the call $\SolveOneEquation{f_2, \{ f_1 \}}$.
	Note that $f_1$ and $f_2$ rewrite 
	respectively as $x (x - 1) (x - 2)=0$ and 
	$x ((2 x - 3) y  - x + 1) = 0$.
	Clearly $x=0$ is a common solution while, 
	for $x=1$ and $x=2$, 
	the equation $f_2$ simply becomes $y=0$ and $y=1$, 
	respectively.
	Therefore, $\SolveOneEquation{f_2, \{ f_1 \}}$ returns 3 components:
	\begin{equation*}
	T_1: \left\{
	\begin{array}{r}
	x = 0 
	\end{array}
	\right., \: \: \: \: 
	T_2: \left\{
	\begin{array}{r}
	y = 0 \\
	x = 1 
	\end{array}
	\right. \: \: {\rm and} \: \: 
	T_3: \left\{
	\begin{array}{r}
	y = 1 \\
	x = 2
	\end{array}
	\right. 
	\end{equation*}
	Next, we need to compute
	$\SolveOneEquation{f_3, T_i}$, for $1 \leq i \leq 3$.
	We observe that $f_3$ satisfies both $T_1$
	and $T_2$, hence the first two calls return
	$T_1$
	and $T_2$. For the third one, we note that $f_3$
	simply becomes $z=0$ at $(y,x)=(1,2)$.
	Finally, $\SolveSystemIncrementally{\{ f_1, f_2, f_3 \}}$
	returns the following 3 components:
	\begin{equation*}
	T_1: \left\{
	\begin{array}{r}
	x = 0 
	\end{array}
	\right., \: \: \: \: 
	T_2: \left\{
	\begin{array}{r}
	y = 0 \\
	x = 1 
	\end{array}
	\right. \: \: {\rm and} \: \: 
	T_3: \left\{
	\begin{array}{r}
	z= 0 \\
	y = 1 \\
	x = 2
	\end{array}
	\right. 
	\end{equation*}
\end{example}

While the previous example shows an opportunity
for parallel computations, namely 
computing $\SolveOneEquation{f_3, T_i}$, for $1 \leq i \leq 3$
in parallel, it hides a major difficulty: possible unbalanced
work. Here is an illustrative example of this latter fact.

\begin{example}
	\label{ex:2}
	Consider the polynomial system:
	\begin{equation*}
	\left\{
	\begin{array}{rl}
	f_1:  & \left( x-1 \right) x =0 \\
	f_2:  &  x \left( {y}^{n}-x \right) + \left( x-1
	\right)  \left( y-x \right) =0 \\
	f_3:  & \left( {y}^{n+1}+1 \right) z-x  =0
	\end{array} \right.
	\end{equation*}
	where $n \geq 2$ is integer.
	Calling $\SolveOneEquation{f_1, \varnothing}$ 
	yields $\{ f_1 \}$, as in Example~\ref{ex:1}.
	Now, $\SolveOneEquation{f_2, \{ f_1 \}}$ returns 3 components
	by means of simple calculations not detailed here:
	\begin{equation*}
	T_1: \left\{
	\begin{array}{r}
	\sum_{i=0}^{n-1} y^i=0 \\
	x = 1 
	\end{array}
	\right., \: \: \: \: 
	T_2: \left\{
	\begin{array}{r}
	y = 1 \\
	x = 1 
	\end{array}
	\right. \: \: {\rm and} \: \: 
	T_3: \left\{
	\begin{array}{r}
	y = 0 \\
	x = 0
	\end{array}
	\right. 
	\end{equation*}
	Next, we compute
	$\SolveOneEquation{f_3, T_i}$, for $1 \leq i \leq 3$.
	Elementary 
	calculations not reported here produce  3 components:
	\begin{equation*}
	T_1: \left\{
	\begin{array}{r}
	(y+1) z + x =0 \\
	\sum_{i=0}^{n-1} y^i=0 \\
	x = 1 
	\end{array}
	\right., \: \: \: \: 
	T_2: \left\{
	\begin{array}{r}
	2 z + 1=0 \\
	y = 1 \\
	x = 1 
	\end{array}
	\right. \: \: {\rm and} \: \: 
	T_3: \left\{
	\begin{array}{r}
	z = 0 \\
	y = 0 \\
	x = 0
	\end{array}
	\right. 
	\end{equation*}
	One should observe  that the first equation in $T_2$ (resp. $T_3$)
	is obtained by evaluating $f_3$ at $(x, y) = (1, 1)$ (resp.  $(0, 0)$),
	which is a trivial operation, the cost of which can be considered
	constant.
	Now, one should notice that the first equation in $T_1$
	is not $f_3$ itself, but $(y+1) z + x =0$.
	This simplification requires dividing
	$y^{n+1}+ 1$ by $\sum_{i=0}^{n-1} y^i$, which, for $n$ large enough,
	is arbitrarily expensive. As a result, 
	the work load in $\SolveOneEquation{f_3, T_i}$, for $1 \leq i \leq 3$,
	can be arbitrarily unbalanced.
\end{example}

Example~\ref{ex:2} shows that parallelizing
the for-loop in Algorithm~\ref{Algo:SolveIncrementally}
may not bring much benefit. Nevertheless,  this does not imply
that, for other examples, it should not be attempted.  In fact, the call
$\SolveOneEquation{p, T}$ itself has various opportunities for 
concurrency. Combined with that Algorithm~\ref{Algo:SolveIncrementally},
many practical examples, but not all, may benefit from 
a careful implementation of concurrency, as explained
in Sections~\ref{sec:practice} and \ref{sec:Implementation}.

\subsection{The Chinese remainder theorem}
\label{sec:CRT}
One of the most fundamental results in algebra is the Chinese remainder
theorem (CRT). Let $m_1$ and $m_2$ be two relatively prime numbers,
so that there exist integers $u_1, u_2$ satisfying $u_1 m_1 +
u_2 m_2 = 1$.
The CRT states that
the residue class ring ${\Z}/(m_1 \, m_2)$ is 
{\em isomorphic} 
to the {\em direct product} of the residue class rings ${\Z}/m_1$ and
${\Z}/m_2$, denoted as ${\Z}/m_1 \otimes {\Z}/m_2$, which
writes:\footnote{Note that for $m\in{\Z}$, we write ${\Z}/m$ as a
	simplified notation for ${\Z}/m{\Z}$.}
\begin{equation}
\label{eq:CRT1}
{\Z}/(m_1 \, m_2) \ \ \equiv \ \ 
{\Z}/m_1 \otimes {\Z}/m_2.
\end{equation}
In the above direct product, 
elements are pairs of residues $(r_1, r_2)$
with $r_i \in {\Z}/m_i$ and operations (addition, subtraction,
multiplication) are performed component-wise.  That is, $(r_1, r_2) +
(s_1, s_2) = (r_1 + s_1, r_2 + s_2)$ and $(r_1, r_2) \times (s_1, s_2)
= (r_1 \times s_1, r_2 \times s_2)$ both hold for all $r_i, s_i \in
{\Z}/m_i$, for $1 \leq i \leq 2$.
The isomorphism stated in Equation (\ref{eq:CRT1})
is a one-to-one map $\Phi$ by which:
\begin{enumerateshort}
	\item the image of 
	$r \in {\Z}/(m_1 \, m_2)$ is $(r \mod{ m_1 }, r \mod{ m_2 })$,
	\item the preimage of $(r_1, r_2) \in {\Z}/m_1 \otimes {\Z}/m_2$
	is $(r_1 m_1 u_1 + r_2 m_2 u_2) \mod{m_1 \, m_2}$, and
	\item the image of a sum (resp. product) is the sum
	(resp. product) of the images, that is,
	$\Phi (r + s) = \Phi (r) + \Phi (s)$ and
	$\Phi (r \times s) = \Phi (r) \times \Phi (s)$.
\end{enumerateshort}
These properties imply that 
\begin{enumerateshort}
	\item any computation
	(i.e. combination of additions, subtractions,
	multiplications)
	in ${\Z}/(m_1 \, m_2)$ can be
	transported to ${\Z}/m_1 \otimes {\Z}/m_2$, where
	\item computations split into two ``independent coordinates'', 
	one in ${\Z}/m_1$ and one in
	${\Z}/m_2$, which can be considered concurrently.
\end{enumerateshort}

The CRT generalizes in many ways. First, with
several pairwise relatively prime integers $m_1, m_2, \ldots,
m_e$, yielding the following:
\begin{center}
	${\Z}/(m_1 m_2 \cdots m_e)
	\ \ \equiv \ \ 
	{\Z}/m_1 \otimes {\Z}/m_2 \otimes \cdots \otimes
	{\Z}/m_e$
\end{center}
Second, with polynomials instead of integers.
For this second case and to be precise, consider
the field ${\Q}$ of rational numbers and the ring
of univariate polynomials in $x$ over ${\Q}$,
denoted by ${\Q}[x]$.
For non-constant polynomials $p_1, p_2, \ldots, p_e$
that are relatively prime (meaning here that
any two polynomials $p_i, p_j$, for $1 \leq i < j \leq e$
have no common factors, except rational numbers) we have
the following isomorphism:\footnote{Note that for a polynomial $p$ we write ${\Q}[x]/p$ as a simplified notation for ${\Q}[x]/\langle p\rangle$.}
\begin{equation}
\label{eq:CRT2}
{{\Q}[x]}/(p_1 p_2 \cdots p_e)
\ \ \equiv \ \
{{\Q}[x]}/p_1 \otimes {{\Q}[x]}/p_2 \otimes \cdots \otimes
{{\Q}[x]}/p_e
\end{equation}
When the polynomials $p_1, p_2, \ldots, p_e$ are irreducible, then
each of ${{\Q}[x]}/p_1$, ${{\Q}[x]}/p_2$, \ldots,
${{\Q}[x]}/p_e$ is in fact a field.  In such case, 
the residue class ring
${{\Q}[x]}/(p_1 p_2 \cdots p_e)$
is a {\em direct product of fields}, an algebraic
structure with  interesting properties.
For instance, if $e=2$, $p_1 = x^2 -2$ and $p_2 = x^2 -3$, then
${{\Q}[x]}/p_1$ and ${{\Q}[x]}/p_2$ are the field
extensions ${\Q}(\sqrt{2})$ and ${\Q}(\sqrt{3})$. Those latter fields
consist of all numbers obtained by adding, multiplying
rational numbers together with $\sqrt{2}$ and $\sqrt{3}$,
respectively.

When the polynomials $p_1, p_2, \ldots, p_e$ are
square-free\footnote{For any field or a direct product of fields
	${\K}$, a univariate non-constant polynomial $p \in {\K}[x]$ is
	square-free whenever for any other non-constant polynomial $q$,
	the square of $q$, that is, $q^2$, does \textit{not} divide $p$.}
then \emph{each} of ${{\Q}[x]}/p_1$, ${{\Q}[x]}/p_2$,
\ldots, ${{\Q}[x]}/p_e$ is in fact a direct product of fields (DPF).
(This claim directly follows from the previous paragraph.)

Now, a third generalization of the CRT is as follows.
For non-constant  pairwise relatively prime polynomials 
$f_1, f_2, \ldots, f_e$ in ${\K}[x]$, where ${\K}$ 
is a DPF, we have the following isomorphism: 
\begin{equation}
\label{eq:CRT2=3}
{{\K}[x]}/(p_1 p_2 \cdots p_e)
\ \ \equiv \ \
{{\K}[x]}/p_1 \otimes {{\K}[x]}/p_2 \otimes \cdots \otimes
{{\K}[x]}/p_e
\end{equation}

\subsection{The D5 principle}
\label{sec:TheD5Principle}

It follows from the discussion above that DPFs generalize
fields. Moreover, square-free polynomials can be used to build
extensions of DPFs.  Hence, one can ``almost'' work with DPFs as if
they were fields.  The ``almost'' comes from the fact that DPFs have
zero-divisors while fields, by definition, do not.  Returning to the
example
\begin{center}
	${{\Q}[x]}/(p_1 p_2)
	\ \ \equiv \ \ 
	{{\Q}[x]}/p_1 \otimes  {{\Q}[x]}/p_2$,
\end{center}
with $p_1 = x^2 -2$ and $p_2 = x^2 -3$, we observe that $p_2$ (as an
element of ${{\Q}[x]}/(p_1 p_2)$) is mapped to $(p_2
\mod{ p_1 }, p_2 \mod{ p_2 }) = (1, 0)$, since $p_2 = 1 + p_1$. Now
observe that $(1, 0) \times (0, 1) = (0,0)$ holds.  Hence $(1, 0),
(0, 1)$ are two non-zero elements of 
${{\Q}[x]}/p_1 \otimes  {{\Q}[x]}/p_2$,
with zero as product. That is, $(1, 0),
(0, 1)$ are \emph{zero-divisors}.

Zero-divisors do not have multiplicative inverses and, thus, are an
obstacle to certain operations.  For instance, consider two univariate
polynomials $a, b \in {\L}[y]$, with $b$ non-constant, over a direct
product of fields ${\L}$. Then, the polynomial division of $a$ by $b$
is uniquely defined provided that the leading coefficient of $b$ is
invertible, that is, not a zero-divisor. Indeed, every non-zero
element in a DPF is either invertible or a zero-divisor.

The {\em D5 principle} \cite{della1985new} states that computations
in ${\L}[y]$ can be performed as if ${\L}$ was a field,
provided that, when a zero-divisor is encountered, 
computations split into separate coordinates, following 
an isomorphism, like the one given
by Equation~(\ref{eq:CRT2}).
We shall explain more precisely what this means.

We assume that
${\L}$ is of the form
${\K}[x] / p$ where $p$  is a square-free non-constant
polynomial and ${\K}$ is either a field (e.g. ${\Q}$)
or another DPF. 
Returning to the polynomial division of $a$ by $b$,
we should decide whether $h$, the leading coefficient
of $b$, is invertible or not. Note that $h$ is a 
polynomial in ${\K}[x]$.
Thus, the invertibility of $h$ modulo $p$ 
can be decided by 
applying the extended Euclidean
algorithm (to which we recursively apply the 
D5 principle, in case ${\K}$ is itself a DPF)
in ${\K}[x]$, with $h$ and $p$ as input.
This produces three polynomials $u, v, g$, with $g \neq 0$, 
such that $u \, h \, + v \, p \ = \ g.$

First, assume that the leading coefficient of  $g$ 
is invertible in ${\K}$. Then, $g$ is a 
greatest common divisor (GCD) of $h$ and $p$.
Two cases arise:
\begin{enumerateshort}
	\item if $g$ is a constant of ${\K}$ (that is, $g$
	has degree zero as a polynomial in ${\K}[x]$)
	then $u g^{-1} h \equiv 1 \, \mod{p}$ holds 
	and we have proved that $h$ is invertible;
	\item if $g$ is not constant (that is,  $g$ has a positive 
	degree in $x$) then, since $p$ is square-free
	resulting in $g$ and $\nicefrac{p}{g}$ being relatively prime,
	CRT gives us,
	\begin{center}
		${\L} \ \ \equiv \ \ {\K}[x] / g \, \otimes {\K}[x] / \frac{p}{g}$,
	\end{center}
\end{enumerateshort}
It follows that either $h$ is invertible or computations split.
In the latter case, computations resume independently on each coordinate.

Second, assume that the leading coefficient of $g$ is not invertible in
${\K}$. Then, reasoning by induction, the DPF ${\K}$ can be replaced
by a direct product so that, here again, computations are performed
independently on each coordinate.

\section{The Triangularize algorithm}
\label{sec:Triangularize}

The {\sf Triangularize} algorithm is a procedure for solving systems of
polynomial equations incrementally.  Hence, it follows the algorithmic
pattern discussed in Section~\ref{sec:solvingincrementally}.
The {\sf Triangularize} algorithm was first proposed
in~\cite{moreno2000triangular} and substantially improved
in~\cite{DBLP:journals/jsc/ChenM12}. 
An extension to systems of
polynomial equations and inequalities is presented
in~\cite{DBLP:journals/jsc/LiMS09}.

The components produced by the {\sf Triangularize} algorithm are
called {\em quasi-components}: they are represented by families of
polynomials with remarkable properties called {\em
	regular chains}. We refer to \cite{DBLP:journals/jsc/ChenM12}
for a formal presentation, while Section~\ref{sec:RegularChains}
describes only the properties relevant to the concurrent
execution of {\sf Triangularize}.

\subsection{Regular chains}
\label{sec:RegularChains}

Let ${\L}$ be a direct product of fields (DPF)
as defined in Section~\ref{sec:TheD5Principle}.
Thus, ${\L}$ is of the form
${\K}[x] / p$ where $p$  is a square-free non-constant
polynomial and ${\K}$ is either a field (e.g. ${\Q}$)
or another DPF.

If follows that any DPF ${\L}$ 
\iflongversion
(as defined in Section~\ref{sec:TheD5Principle})
\fi
can be given by a sequence of multivariate polynomials
$T = p_1(x_1), p_2(x_1, x_2), \ldots, p_n(x_1, x_2, \ldots, x_n)$
with coefficients in a field ${\L}_0$, 
such that, for $1 \leq i \leq n$, we have:
\begin{enumerateshort}
	\item $p_i$ is a square-free non-constant
	polynomial in ${\L}_{i-1}[x_i]$,
	\item the leading
	coefficient $h_i$ of $p_i$ is \textit{regular with respect to} ${\L}_{i-1}$, i.e. invertible in ${\L}_{i-1}$, and
	\item ${\L}_{i} := {\L}_{i-1}[x_i] / p_i$ and ${\L}_n = {\L}$.
\end{enumerateshort}
Two cases are of practical importance.
First, if ${\L}_0 = {\Q}$, then any 
point $(z_1, z_2, \ldots, z_n)$ with complex number coordinates
satisfying 
\begin{center}
	$p_1(z_1) = p_2(z_1, z_2) = \cdots = p_n(z_1, z_2, \ldots, z_n) = 0$,
\end{center}
is called a {\em solution} of $T$; the set of those solutions
is denoted by $V(T)$: this is a finite set and $T$ is
called {\em zero-dimensional}.
Second, if ${\L}_0 $ is a field of rational 
functions\footnote{A rational function is a fraction
	of two polynomials, for instance $\frac{2t}{1 + t^2}$.}
with variables $t_1, \ldots, t_d$, then any point
$(t_1, \ldots, t_d, z_1, \ldots, z_n)$ with complex number coordinates
satisfying for $1 \leq i \leq n$:
\begin{center}
	$p_i(t_1, \ldots, t_d, z_1, \ldots, z_i) = 0
	\ {\rm and} \  h_i(t_1, \ldots, t_d, z_1, \ldots, z_{i-1}) \neq 0$,
\end{center}
is called a {\em solution} of $T$; the set of those solutions
is denoted by $W(T)$ and called a {\em quasi-component}: 
this set has {\em dimension} $d$.
In either case, the sequence $T$ of polynomials is called 
a {\em regular chain}.

\subsection{The main procedure and its core routines}
\label{sec:mainProcedure}

Algorithm~\ref{Algo:Triangularize} states the top-level
procedure of the {\sf Triangularize} algorithm.
Let us denote by $f_1, f_2, \ldots, f_m$ the polynomials in $F$ and by
$x_1, x_2, \ldots, x_n$ their variables.  
Clearly, computations are essentially done
by the core routine {\sf Intersect}, which corresponds to 
the {\sf SolveOneEquation} routine from 
Algorithm~\ref{Algo:SolveIncrementally}; naturally, {\sf Triangularize}
itself corresponds to {\sf SolveSystemIncrementally}.
For a polynomial $p \in {\Q}[x_1, x_2, \ldots, x_n]$
and a regular chain $T$ given by polynomials from
${\Q}[x_1, x_2, \ldots, x_n]$, the function call
$\Intersect{p, T}$ returns regular chains
$T_1, T_2, \ldots, T_e$ so that
the union $W(T_1) \cup W(T_2) \cup \cdots \cup W(T_e)$ is
an approximation\footnote{To be precise, if
	$\overline{W(T)}$ is the topological closure of $W(T)$,
	in the sense of the Zariski topology, then
	$W(T) \cap V(p) \subset 
	W(T_1) \cup W(T_2) \cup \cdots \cup W(T_e) \subset
	\overline{W(T)}  \cap V(p).$}
of the intersection $W(T) \cap V(p)$.
This approximation is, in fact, very sharp
and this is why $\Triangularize{F}$ is capable of producing 
regular chains such that the union of their zero sets is 
exactly equal to the zero set of $F$, denoted $V(F)$.

\begin{algorithm}[htb]
	\caption{\Triangularize{F}\label{Algo:Triangularize}}
	\KwIn{a finite set of polynomials $F$}
	\KwOut{a set of regular chains such that the union of their zero sets is $V(F)$}
	{\bf if} $|F| == 0$ {\bf then} \textbf{return} $[\{\}]$\;
	Choose a polynomial $p\in F$\;
	\For{$T {\bf\ in\ } \Triangularize{F\setminus\{p\}}$}{
		{\bf yield} $\Intersect{p, T}$\;
	}
\end{algorithm}

\begin{figure}[htb!]
	\centering
	\usetikzlibrary{backgrounds}
	\begin{tikzpicture}[nodestyle/.style={font=\sffamily},
	]
	\node[nodestyle] (tri) at (0,3.8) {Triangularize};
	\node[nodestyle] (rrc) at (2,3) {RRC};
	\node[nodestyle] (int) at (0,3) {Intersect};
	\node[nodestyle] (intfree) at (-2.5,2.5) {IntersectFree};
	\node[nodestyle] (clean) at (0,2.0) {CleanChain};
	\node[nodestyle] (intalg) at (2.7,1.5) {IntersectAlgebraic};
	\node[nodestyle] (reg) at (-2.2,1.5) {Regularize};
	\node[nodestyle] (reggcd) at (0, 1,0) {RegularGCD};
	\node[nodestyle] (ext) at (-3.2,0.75) {Extend};

	\draw[->, >=stealth] ([xshift=-0.3em]ext.south) to[out=-135, in=180, looseness=1.5] ([yshift=-1em]ext.south) to[out=0, in=-45, looseness=1.5] ([xshift=0.3em]ext.south);
	\draw[->, >=stealth] ([xshift=-0.2em]reg.south) to[out=-135, in=180, looseness=1.4] ([xshift=0.1em,yshift=-1em]reg.south) to[out=0, in=-45, looseness=1.5] ([xshift=0.4em]reg.south);
	\draw[->, >=stealth] ([xshift=-0.3em]intalg.south) to[out=-135, in=180, looseness=1.5] ([yshift=-1em]intalg.south) to[out=0, in=-45, looseness=1.5] ([xshift=0.3em]intalg.south);
	
	\draw[->, >=stealth] (tri.south) -- (int.north);
	\draw[->, >=stealth] (tri.south east) -- (rrc.north west);
	\draw[<->, >=stealth] ([yshift=-0.1em,xshift=-0.05em]int.west) -- (intfree.north east);
	\draw[<->, >=stealth] ([yshift=-0.1em,xshift=0.05em]int.east) -- (intalg.north);
	\draw[->, >=stealth] (reg.north) -- ([xshift=-1em]int.south);

	\draw[<->, >=stealth] ([xshift=0.2em]ext.north) -- ([yshift=0.2em,xshift=0.3em]reg.south west);
	\draw[<->, >=stealth] ([yshift=0.1em]reggcd.west) -- ([yshift=0.2em]reg.south east);
	\draw[<-, >=stealth] ([yshift=0.1em]reggcd.east) -- ([yshift=0.2em]intalg.south west);
	\draw[->, >=stealth] ([yshift=-0.2em]intalg.north west) -- ([yshift=-0.2em]clean.east);
	\draw[->, >=stealth] (intfree.south) -- ([yshift=-1.7em]intfree.south);
	\draw[->, >=stealth] (int.south) -- (clean.north);
	\draw[->, >=stealth] ([yshift=-0.2em]clean.west) -- ([yshift=-0.2em]reg.north east);
	
	\end{tikzpicture}
	\caption{The flow graph of the Triangularize algorithm.}\label{fig:triangularizeflowgraph}
\end{figure}
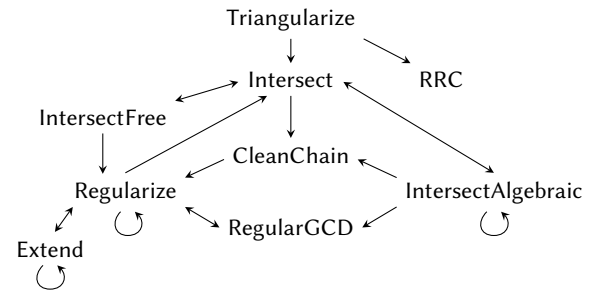

During the execution of $\Intersect{p, T}$, two mechanisms
lead computations to split into different cases:
\begin{enumerateshort}
	\item case distinctions of the form
	``either $a = 0$ or $a \neq 0$'', where $a$ is a polynomial
	by which one wants to divide another polynomial,
	like in $a x = b$ when solving for $x$.
	\item Case distinctions resulting from the application
	of the D5 principle, as described in Section~\ref{sec:TheD5Principle}.
\end{enumerateshort}
Case distinctions of this latter type are generated
by {\sf Regularize}, a core routine of
{\sf Triangularize}. Figure~\ref{fig:triangularizeflowgraph},
taken from~\cite{DBLP:journals/jsc/ChenM12}, 
displays the ``depends on'' graph of the main routines
of {\sf Triangularize}.

\subsection{Removal of the redundant components}
\label{sec:RedundantComponents}

As we have seen with Examples~\ref{ex:1} and \ref{ex:2}, computing
$\Intersect{p, T}$ often leads to case distinctions of the form:
either $a = 0$ or $a \neq 0$.
Those are likely to produce components that are redundant,
while not being incorrect.
Here's a trivial example.
Consider $F := \{ f_1, f_2 \}$ with $f_1 : y x + y = 0$
and $f_2 : y = 0$.
\Intersect{f_1, \varnothing} yields two regular chains
$T_1, T_2$ with $T_1 : y=0$ and $T_2 : x+1 = 0$.
Then, \Intersect{f_2, T_1} yields $T_1$ itself
while \Intersect{f_2, T_2} yields $T_3$, with $T_3 : x+1 = 0, y=0$.
Clearly, $T_3$ is a special case of $T_1$, that is,
we have $W(T_3) \subseteq W(T_1)$ and $T_3$ is redundant.

Precisely, letting $T_1, T_2, \ldots, T_e$ be the output regular
chains of {\Triangularize{F} in Algorithm~\ref{Algo:Triangularize},
	regular chain $T_i$ is {\em redundant}
	when:
	\begin{center}
		$W(T_i) \subseteq W(T_1)
		\cup \cdots \cup W(T_{i-1}) \cup W(T_{i+1}) \cup \cdots \cup W(T_e)$.
	\end{center}
	
	There exist criteria and algorithmic procedures for testing whether a
	regular chain is redundant or not.  However, applying those criteria and
	procedures can be very computationally expensive. 
	As a result, one should carefully consider when to attempt performing
	those computations.  Two strategies are meaningful here:
	\begin{description}
		\item[{\bf barrier}:] test for redundant
		components on every output returned by a call to {\sf Triangularize},
		including recursive calls;
		\item[{\bf barrier-free}:] test for redundant
		components only on the final output returned by \Triangularize{F}.
	\end{description}
	
	With the barrier strategy, all calls of the 
	form \Intersect{f_i, T} must complete before
	starting to execute the calls of the 
	form \Intersect{f_{i+1}, T'}, for $1 \leq i \leq m$,
	where $F := \{ f_1, f_2, \ldots, f_m \}$ is the input
	of {\sf Triangularize}.
	For this reason, this strategy computes the solution
	{\em level-by-level}, hence we call it {\em Level} in subsequent sections.
	With the barrier-free approach, 
	particularly when using parallelization, some calls
	of the form \Intersect{f_{i+1}, T'}
	may be executed before completing all calls
	of the form \Intersect{f_{i}, T'}.
	We call this method {\em Bubble} in subsequent
	sections because it allows components to 
	``bubble up'' to higher levels of the recursion in
	{\sf Triangularize}.
	Section~\ref{sec:practice} considers the
	parallelization and implementation details of these two strategies
	and redundant component removal.

	\subsection{Computing generic solutions}
	\label{sec:genericsols}
	
	Another major feature of the {\sf Triangularize} algorithm leads to
	different strategies and should be carefully considered towards
	performance in both serial and concurrent execution.
	This feature deals with the ``scope'' of the solution set.
	To be precise, it is possible to limit the computations
	of {\sf Triangularize} to the so-called 
	{\em generic solutions}.\footnote{Denoting by $T_1, T_2, \ldots, T_e$ the output regular
		chains of \Triangularize{F} in Algorithm~\ref{Algo:Triangularize},
		a regular chain $T_i$ is {\em generic}
		whenever we have:
		$W(T_i) \not\subseteq \overline{W(T_1)}
		\cup \cdots \cup \overline{W(T_{i-1})} \cup \overline{W(T_{i+1})} \
		\cup \cdots \cup \overline{W(T_e)}$.}
	We refer again to \cite{DBLP:journals/jsc/ChenM12} for a formal
	presentation and we shall simply use an illustrative example here.
	Consider the input system $F$ with the unique equation $a x = b$
	with the intention of solving for $x$.
	\Triangularize{F} returns two regular chains $T_1: a x = b$
	and $T_2: a=0, b=0$. The first one encodes all solutions
	for which $a \neq 0$ holds.
	``Generically'' (say, by randomly picking a value of $a$)
	we have $a \neq 0$. Hence $T_1$ encodes all generic solutions
	while $T_2$ correspond to the non-generic ones, also
	called {\em degenerate solutions}.
	
	The {\sf Triangularize} algorithm can be executed in two modes,
	the names of which are from mathematicians who have
	studied strategies corresponding to those modes:
	\begin{description}
		\item[{\bf Kalkbrener}:] 
		non-generic solutions are likely not computed,
		\item[{\bf Lazard-Wu}:] 
		all solutions (generic or not) are computed.
	\end{description}
	The former mode is derived from the latter by a tree pruning
	technique. 
	To clarify this, we can understand the {\sf Triangularize} algorithm to
	be performing a breadth-first search (BFS) of a tree of intersections of 
	polynomials and regular chains. See Figure~\ref{fig:tritree1} for an 
	example of this. The Lazard-Wu decomposition finds 
	all paths of the tree that yield solutions of the input system. The
	Kalkbrener decomposition, however, cuts off branches of the tree 
	for which the \emph{height} of a regular chain, i.e., the number of polynomials 
	in the chain, exceeds a certain bound.\footnote{Specifically, this bound is given 
		by the number $m$ of input polynomials in $F$. That the height of chains can be used to prune branches
		follows from an application of {\em Krull's height theorem} 
		(see~\cite{DBLP:journals/jsc/ChenM12}, for details).} In this way, paths of the tree 
	that exceed the height bound are ``pruned''. In the 
	analogy to BFS, Kalkbrener decomposition is similar to
	Dijkstra's algorithm for finding optimal paths
	(where height is analogous to edge weight).
	For polynomial systems that are 
	computationally expensive to solve, it is often the  case
	that the Kalkbrener mode can use much less computing
	resources (CPU time, memory) than the Lazard-Wu mode.
	
	\begin{figure}[htb]
		\includegraphics[width=0.48\textwidth, trim={25px, 72px, 50px, 90px},clip]{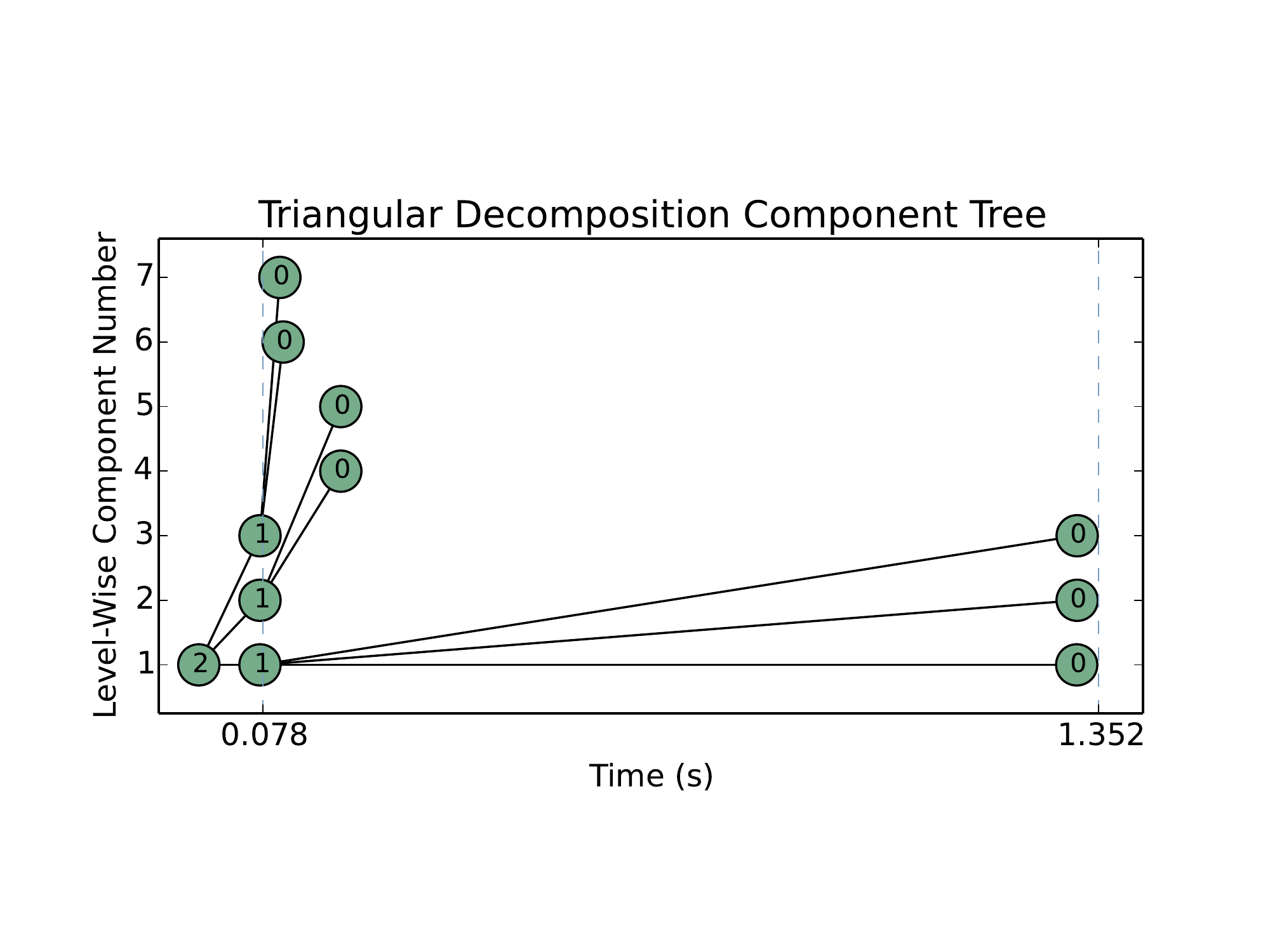}
		\caption{A tree of components from the splitting of computations
			through the \textsf{TriangularizeLevel} algorithm run on a system of 2 polynomials in 2 variables.
			The label on each node represents the component's dimension while the grid lines indicate the time at which each level was complete.}\label{fig:tritree1}
	\end{figure}

	%

\section{Exploiting Concurrency in Triangular Decomposition}
\label{sec:practice}

The {\sf Triangularize} algorithm is highly complex, both in
time-complexity and in conceptual complexity.  This can seen from the
flow graph in Figure~\ref{fig:triangularizeflowgraph} above.
While all of the subroutines are greatly important to the algorithm we
do not detail them here; for a more complete description see
\cite{DBLP:journals/jsc/ChenM12}.  Here we illustrate how {\sf
	Triangularize} and its subroutines hold opportunities for
parallelism from an algorithmic point of view.

We begin in Section~\ref{sec:gen-forkjoin} by discussing the
concurrency paradigms applicable to the {\sf Triangularize}
algorithm.  The parallelization of the removal of redundant components
and the {\sf Triangularize} algorithm itself are then discussed in
Sections~\ref{sec:rrc} and \ref{sec:bubblevslevel}, respectively.

\subsection{The fork-join model, generators, and the producer-consumer pattern}
\label{sec:gen-forkjoin}

There are many patterns common to parallel processing, algorithms, and
programming; fork-join, map-reduce, and the stencil pattern are
notable.  Fork-join is an idiomatic pattern for parallel processing at
the heart of many concurrency platforms such as \cilk, \openmp, and
Intel \textit{Threading Building Blocks} (TBB)
\cite{mccool2012structured}.

The ubiquitous divide-and-conquer (DnC) algorithm design readily
admits itself to parallelism using the fork-join model.
Briefly, the \textit{fork} divides work---and the program control flow---into 
multiple independent streams, each run in parallel, 
and then \textit{joins} those streams at a synchronization 
point, before proceeding again serially.
DnC, with its multiple recursive calls, 
lends itself to the fork-join model where the recursive calls can be forked
and done in parallel.

Another common pattern in parallel processing---although rarely
labelled directly as such---is generators. 
Indeed, generators are a more general term from computer programming 
to mean simply a special routine which can yield a sequence of values
one a time rather than as a complete list. 
Consider the more common (in parallel processing) 
producer-consumer pattern \cite{mccool2012structured};
in truth, this can be thought of as an \textit{asynchronous generator}
where yields in the generator are instead the producer pushing data to the shared queue. 
By implementing generators in a producer-consumer scheme
concurrency is easy to achieve where both the generator (producer),
and the caller to the generator (consumer), may run concurrently. 

The next two sections detail our use of the fork-join model and
these asynchronous generators
to exploit possible parallelism. 

\subsection{Removing redundant components}
\label{sec:rrc}
To remove redundant components efficiently, we must address two
issues: how to efficiently determine single inclusions,
$W(T_i)\stackrel{?}{\subseteq}W(T_j)$; 
and how to efficiently remove
redundant components from a large set. Both these issues are addressed
in \cite{xie2007fast}.  The first issue is addressed by using a
heuristic algorithm for the inclusion test \textsf{IsNotIncluded},
which is very effective (removes most redundant
components) in practice \cite[see][pp. 166--169]{xie2007fast}.  
The large numbers of comparisons from a large
set are handled efficiently by structuring the computation as a
divide-and-conquer algorithm.

Given a set $\mathcal{T}=\{T_1,\ldots,T_e\}$ of regular chains, Algorithm \ref{Algo:RemoveRedundantComponents}, 
{\sf RemoveRedundantComponents}($\mathcal{T}$), abbreviated {\sf RRC}($\mathcal{T}$),
removes redundant chains by dividing $\mathcal{T}$ into
two subsets of approximately equal size and recursively calling {\sf RRC}
on these subsets 
to produce two irredunant sets of chains. These recursive calls are computed
in parallel using the fork-join pattern.
A second routine {\sf MergeIrredundantLists} is
then called, which, as the recursion unwinds, merges these sets into a single irredundant set. 
The base case of this recursion is a set of size 1, where no redundancies
are possible.

The routine  {\sf MergeIrredundantLists}, given as Algorithm \ref{Algo:MergeIrredundantLists},
works by first removing all chains from $\mathcal{T}_1$ that are included in
some chain of $\mathcal{T}_2$ to produce a (possibly) smaller list $\mathcal{T}_1'$. Then, all chains
in $\mathcal{T}_2$ that are contained in $\mathcal{T}_1'$ are removed.
Removal in each direction is accomplished by two nested loops, with the 
actual inclusion test performed by another routine {\sf IsNotIncluded}$(T_1,T_2)$, which returns
{\bf true} is $T_1$ is definitely not included in $T_2$ and {\bf false} otherwise (inclusion does 
hold or inclusion relation cannot be determined). 

Each for loop in \textsf{MergeIrredudantSets} 
is embarrassingly parallel, since each comparison test is independent.
Hence, we make the outer for loop a {\bf parallel\_for} to take advantage of this.
Though one could also parallelize the inner for loop, we find the amount of work 
at this lower level does not warrant the overhead of parallelism.

\begin{algorithm}[htb]
	\caption{{\sf RemoveRedundantComponents}$(\mathcal{T})$\label{Algo:RemoveRedundantComponents}}
	\KwIn{a finite list $\mathcal{T}=[T_1,\ldots,T_e]$ of regular chains}
	\KwOut{a list $\mathcal{T}'$ of regular chains}
	\lIf {$e == 1$} {{\bf return} $[T_1]$}
	$\ell = \lceil e/2 \rceil$\;
	$\mathcal{T}_{\leq \ell}=[T_1,\ldots,T_{\ell}]$; $\mathcal{T}_{> \ell}=[T_{\ell+1},\ldots,T_{e}]$\;
	{\bf spawn} $\mathcal{T}_1$ = {\sf RemoveRedundantComponents}($\mathcal{T}_{\leq \ell}$)\;
	$\mathcal{T}_2$ = {\sf RemoveRedundantComponents}($\mathcal{T}_{>\ell}$)\;
	{\bf sync}\;
	$\mathcal{T}' =\ ${\sf MergeIrredundantLists}($\mathcal{T}_1,\mathcal{T}_1$)\;
	\textbf{return} $\mathcal{T}'$\;
\end{algorithm}

\begin{algorithm}[htb]
	\caption{{\sf MergeIrredundantLists}($\mathcal{T}_1,\mathcal{T}_2)$\label{Algo:MergeIrredundantLists}}
	\KwIn{two finite lists $\mathcal{T}_1$ and $\mathcal{T}_2$ of regular chains}
	\KwOut{a list of regular chains}
	$\mathcal{T}_1' = [\,]$; $\mathcal{T}_2' = [\,]$\;
	\ParFor{$T_2$ {\bf in} $\mathcal{T}_2$}{
		\For{$T_1$ {\bf in} $\mathcal{T}_1$}{
			\lIf {{\sf IsNotIncluded}$(T_1,T_2)$}{ $\mathcal{T}_1' \text{\texttt{ += }} T_1$}
		}
	}
	\ParFor{$T_1'$ {\bf in} $\mathcal{T}_1'$}{
		\For{$T_2$ {\bf in} $\mathcal{T}_2$}{
			\lIf {{\sf IsNotIncluded}$(T_2,T_1')$} {$\mathcal{T}_2'\text{\texttt{ += }} T_2$}
		}
	}
	\textbf{return} $\mathcal{T}_2'$\;
\end{algorithm}

\subsection{Triangularize: Bubble vs. Level}
\label{sec:bubblevslevel}

\begin{algorithm}[htb]
	\caption{\TriangularizeBubble{F}\label{Algo:TriangularizeBubble}}
	\KwIn{a finite set of polynomials $F$}
	\KwOut{a set of regular chains such that the union of their zero sets is $V(F)$}
	$\mathcal{T} = [\,]$\;
	\For{$T \text{\textnormal{\textbf{ in }}} \Triangularize{F}$}{
		$\mathcal{T} \text{\texttt{ += }} T$\;	
	}
	\textbf{return} $\RemoveRedundantComponents{\mathcal{T}}$\;
\end{algorithm}

\begin{algorithm}[htb]
	\caption{\TriangularizeLevel{$F$}\label{Algo:TriangularizeLevel}}
	\KwIn{a finite set of polynomials $F=\{f_1,\ldots,f_m\}$}
	\KwOut{a set of regular chains such that the union of their zero sets is $V(F)$}
	
	$\mathcal{T} = [\{\}]$\;
	\For{$i = 1\,..\,m$}{
		$e = |\mathcal{T}|$; $\mathcal{T}' = [\,]$\;
		\ParFor{$i = 1\,..\,e$} {
			$\mathcal{T}' \text{\texttt{ += }} \Intersect{f_i, \mathcal{T}[i]}$\;
		}
		$\RemoveRedundantComponents{\mathcal{T}',\mathcal{T}}$\;
	}
	\textbf{return} $\mathcal{T}$\;
\end{algorithm}

We saw in Section~\ref{sec:RedundantComponents} that the decision of
\emph{when} to remove redundant components gives rise to two different 
versions of the {\sf Triangularize} algorithm. We now discuss these algorithmic 
variants in more detail and describe their opportunities for concurrent computations.

The barrier-free Bubble strategy, given as Algorithm~\ref{Algo:TriangularizeBubble},
puts off the removal of redundant components until
after the top level call to {\sf Intersect} in Algorithm~\ref{Algo:Triangularize}
has returned. The advantage of this approach for concurrency is that using
asynchronous generators allows {\sf Intersect} to yield components to the next higher level of
{\sf Triangularize} in the recursion stack, and thus allowing subsequent {\sf Intersect} tasks to be run
as early as possible.
We only call {\sf RRC} when
the top level call to
{\sf Triangularize} has returned all components.

The barrier-using Level strategy, on the other hand, recognizes opportunities
for speed-up by avoiding the repetition of expensive computations 
that redundant components can cause. 
The algorithm, given as Algorithm~\ref{Algo:TriangularizeLevel}, 
takes advantage of the fact 
that the solution can be computed incrementally, intersecting one polynomial at a 
time with the components of the previous solution, and removes redundancies after 
each round of intersections. 
This is a barrier strategy because there is an enforced 
synchronization point at the end of each recursive call to {\sf Triangularize} 
in Algorithm~\ref{Algo:Triangularize}. 
Algorithm~\ref{Algo:TriangularizeLevel} realizes this by converting the 
recursion to an iteration.

The opportunities for concurrent computing are different in this case, tracing 
to the fact that each intersection of a polynomial and a component is independent.
Accordingly, Algorithm~\ref{Algo:TriangularizeLevel} uses a {\bf parallel\_for} to 
execute the {\sf Intersect} calls in parallel.

Both of these concurrency techniques describe our
so-called ``coarse-grained'' parallelism.
However, there is also the possibility for
``fine-grained'' parallelism in the subroutines 
of \textsf{Intersect}.
%
As computations split within the subroutines, namely by \textsf{Regularize}
(see Figure~\ref{fig:triangularizeflowgraph}),
each subroutine must pass data (i.e. components) as lists,
essentially creating arbitrary synchronization points as the list accumulates.
Much like the Bubble strategy, 
asynchronous generators and the producer-consumer
scheme
can be used
to \textit{stream} or \textit{pipe} data between subroutines,
allowing data to flow throughout the subroutines, 
and back to the \textsf{Triangularize} algorithm as quickly
as possible in support of further coarse-grained parallelism.

\section{Implementation}
\label{sec:Implementation}

Our triangular decomposition algorithms are implemented 
in the Basic Polynomial Algebra Subprograms (\textsc{BPAS}) library \cite{ chen2014basic}.
The \textsc{BPAS} library is a free and open-source library
for high performance polynomial operations including arithmetic,
real root isolation,
and now, polynomial system solving.
The library is mainly written in C for performance, with a C++
wrapper interface for portability, object-oriented programming, and end-user usability.
Parallelism is already employed in \textsc{BPAS} in its implementations
of real root isolation \cite{CMX11}, dense polynomial arithmetic \cite{DBLP:journals/corr/ChenCMM0X16},
and FFT-based arithmetic for prime fields \cite{covanovMMW2019bigprime}.
These implementations make use of the \cilk\ extension of C/\cpp\
for parallelism.

Our implementation of triangular decompositions follows that of \cite{DBLP:journals/jsc/ChenM12}
and of the \textit{RegularChains} package
of \maple. However, where the \textit{RegularChains} package
is written in the \maple\ scripting language, our implementation is written
mainly in C, like many of our operations,
so that we may finely control memory, data structures, and cache complexity
to obtain better performance 
on modern computer architectures.

The parallelism exploited within our triangular decomposition algorithm
(see Sections~\ref{sec:rrc} and \ref{sec:bubblevslevel}, and Algorithms~\ref{Algo:RemoveRedundantComponents}, 
\ref{Algo:MergeIrredundantLists}, \ref{Algo:TriangularizeBubble},
and \ref{Algo:TriangularizeLevel}) is realized
by a mix of \cilk\ and {\cpp}11 threads.
The divide-and-conquer approach of \textsf{RRC}
is easily parallelized directly using \texttt{cilk\_spawn}
and \texttt{cilk\_sync} to implement the fork-join pattern.
As for \textsf{MergeIrredundantLists}, the nested iteration
over both lists is embarrassingly parallel. 
We parallelize the outer loop with a simple \texttt{cilk\_for} loop. 
Thanks to the efficient heuristic algorithm implementing 
\textsf{IsNotIncluded}, the inner loop presents little work
to justify the overhead of parallelism.

%
Although it would be natural to use a \texttt{cilk\_for} loop to implement the {\bf parallel\_for} in
\textsf{TriangularizeLevel} (Algorithm~\ref{Algo:TriangularizeLevel}),
we noticed that a \texttt{cilk\_for} loop
did not allow for a \textit{grain size} of 1, i.e., 
to allow one thread per loop iteration. 
Hence, for this 
coarse-grained parallelism we instead manually spawn $|\mathcal{T}|-1$ threads,
which each call \textsf{Intersect} on one component, while the main thread
makes the final call to \textsf{Intersect}.
For the coarse-grained parallelism of \textsf{TriangularizeBubble}
(Algorithm~\ref{Algo:TriangularizeBubble}), as well as the 
fine-grained parallelism used in both variations, we make use 
of custom asynchronous generators to implement an asynchronous producer-consumer pattern.
%

\subsection{Asynchronous generators in {\cpp}11}
\label{sec:fineparallel}

Our implementation of asynchronous generators (\texttt{AsyncGenerator}) 
makes use of the standard {\cpp}11 \textit{thread support library}
to achieve an asynchronous producer-consumer pattern. 
This library includes \texttt{std::thread}---a wrapper of \texttt{pthread} on linux---and
synchronization primitives like mutex and condition variables. 
We also make use of the standard {\cpp}11 \textit{function object library} which
attempts to make functions first-class objects in \cpp.

\begin{figure}[htb]
	\begin{lstlisting}[style=mystyle,caption={The \texttt{AsyncGenerator} interface which implements an asynchronous producer-consumer pattern.}, label={lst:asyncgen}]
	template <class Object>
	class AsyncGenerator {
	
	/* Create a generator from a function call. */
	template<class Function, class... Args>
	AsyncGenerator(Function&& f, Args&&... args);
	
	/* Add a new object to the generator. */
	virtual void generateObject(Object& obj) = 0;
	
	/* Finalize the AsyncGenerator by declaring it has
	finished generating all possible objects. */
	virtual void setComplete() = 0;
	
	/* Obtain the next generated Object, returning by reference.
	returns false iff no more objects were available. */
	virtual bool getNextObject(Object& obj) = 0;
	};
	\end{lstlisting}
	\begin{lstlisting}[style=mystyle,caption={The general usage of \texttt{AsyncGenerator} in a producer-consumer scheme.}, label={lst:asyncgenusage}]
	void Regularize(Poly& p, RC& T, AsyncGen<Poly,RC>& gen) {
	/* ... */ 
	gen.generateObject(next);
	}
	
	void IntersectFree(Poly& p, RC& T, AsyncGen<RC>& gen) {
	AsyncGen<Poly,RC> regularRes(Regularize, p.initial(), T);
	Poly,RC next;
	while (regularRes.getNextObject(next)) {  /* ... */   }
	}
	\end{lstlisting}
\end{figure}

Listing~\ref{lst:asyncgen} shows the simple interface of \texttt{AsyncGenerator} which is shared
by both producer and consumer; it is templated by the \texttt{Object} to pass from producer
to consumer.
The general usage pattern of the \texttt{AsyncGenerator} is as follows:
\begin{enumerateshort}
	\item the consumer constructs an \texttt{AsyncGenerator} using
	a function, and its arguments, to execute (the producer);
	\item the \texttt{AsyncGenerator}, on construction, inserts itself into the function arguments;
	\item the \texttt{AsyncGenerator} (possibly) spawns a thread on which to call the function;
	\item the producer function produces output via its parameter \texttt{AsyncGenerator} and its
	\texttt{generateObject} method;
	\item the consumer function uses the
	\texttt{getNextObject} method to consume objects.
\end{enumerateshort}
The pseudo-code in Listing~\ref{lst:asyncgenusage} shows this usage pattern in a 
producer-consumer scheme between \textsf{IntersectFree} and \textsf{Regularize}.
Note that, since we wish to \textit{stream} data between every subroutine of \textsf{Triangularize},
every subroutine is both a producer and a consumer.

We say that an \texttt{AsyncGenerator} may only possibly
spawn a new thread due to some run-time optimizations. 
Where \texttt{AsyncGenerator} is used for 
the coarse-grained parallelism
of \textsf{TriangularizeBubble} it always spawns a new thread
to achieve good task-level parallelism. 
However, due to the highly recursive nature of \textsf{Triangularize}'s subroutines
(see Figure~\ref{fig:triangularizeflowgraph})
it would be unwise to always spawn a new thread. 
Hence, for the subroutines we use a
static thread pool of \texttt{FunctionExecutorThread}s,
rather than spawning new threads.
These threads are initialized before the algorithm and run an event loop,
waiting for functions to be passed to them to be executed. 

The \texttt{AsyncGenerator} is flexible
to the thread pool's shared usage in that it will call the function 
on the current thread if the pool is ever empty. 
This avoids shifting data to different threads for method calls 
which are deep in the call stacks, but does so at higher levels of the recursion
where the pool is not yet empty and the parallelism is coarser. 
Despite this optimization, we still notice that
the overhead of generators can sometimes negatively affect performance, 
particularly in the case of \textsf{TriangularizeLevel}.
We discuss such performance aspects in the next section.

\section{Experimentation \& Discussion}
\label{sec:Experimentation}

\begin{table*}[htb]
	\footnotesize
	\renewcommand{\arraystretch}{1.01}
	\begin{subtable}{0.48\textwidth}
		\centering
		\begin{tabular}{|c||r|c|c||r|c|c|}
			\hline
			& \multicolumn{3}{|c||}{Level} & \multicolumn{3}{|c|}{Bubble} \\
			\cline{2-4}\cline{5-7}
			System Name & S. Time & C & C+F & Time & C & C+F \\
			\hline
			\scriptsize{8-3-config-Li} & 39.267 & 1.78 & 1.79 & 43.256 & 1.35 & 1.35 \\
			\scriptsize{Hairer-2-BGK} & 13.763 & 2.24 & 2.27 & 13.481 & 1.75 & 1.73 \\
			\scriptsize{John5} & 69.622 & 1.08 & 1.08 & 68.222 & 1.05 & 1.04 \\
			\scriptsize{Lazard-ascm2001} & 17.247 & 1.56 & 1.51 & 18.212 & 1.23 & 1.23 \\
			\scriptsize{Liu-Lorenz} & 25.598 & 1.44 & 1.42 & 25.208 & 1.37 & 1.36 \\
			\scriptsize{Mehta2} & 16.314 & 1.54 & 1.49 & 15.793 & 1.11 & 1.11 \\
			\scriptsize{Morgenstein} & 11.909 & 1.52 & 1.53 & 12.550 & 0.99 & 0.98 \\
			\scriptsize{Reif} & 74.327 & 2.42 & 2.44 & 346.986 & 3.19 & 3.18 \\
			\scriptsize{Xia} & 8.110 & 1.3 & 1.31 & 8.177 & 1.1 & 1.1 \\
			\scriptsize{Sys2161} & 6.737 & 2.03 & 2.0 & 26.266 & 3.41 & 3.54 \\
			\scriptsize{Sys2880} & 12.036 & 2.01 & 1.96 & 77.805 & 2.72 & 2.72 \\
			\scriptsize{Sys3054} & 13.684 & 2.21 & 2.25 & 13.622 & 1.75 & 1.78 \\
			\scriptsize{Sys3064} & 17.140 & 1.55 & 1.59 & 18.072 & 1.21 & 1.21 \\
			\scriptsize{Sys3068} & 25.985 & 1.5 & 1.46 & 25.206 & 1.31 & 1.36 \\
			\scriptsize{Sys3073} & 11.952 & 1.53 & 1.54 & 12.552 & 0.99 & 0.98 \\
			\scriptsize{Sys3170} & 8.149 & 1.3 & 1.27 & 8.171 & 1.08 & 1.08 \\
			\scriptsize{Sys3185} & 16.852 & 1.67 & 1.7 & 15.932 & 1.11 & 1.11 \\
			\scriptsize{Sys3238} & 13.697 & 2.23 & 2.25 & 13.545 & 1.76 & 1.77 \\
			\scriptsize{Sys3258} & 38.806 & 1.44 & 1.5 & 45.062 & 1.19 & 1.19 \\
			\scriptsize{Sys3306} & 17.239 & 1.57 & 1.56 & 18.149 & 1.22 & 1.21 \\
			\hline
		\end{tabular}
		\caption{Lazard-Wu}\label{table:lazexamples}
	\end{subtable}
	\hfill
	%
	\begin{subtable}{0.48\textwidth}
		\centering
		\begin{tabular}{|c||r|c|c||r|c|c|}
			\hline
			& \multicolumn{3}{|c||}{Level} & \multicolumn{3}{|c|}{Bubble} \\
			\cline{2-4}\cline{5-7}
			System Name & S. Time & C & C+F & Time & C & C+F \\
			\hline
			\scriptsize{8-3-config-Li} & 24.323 & 1.49 & 1.49 & 24.153 & 1.06 & 1.39 \\
			\scriptsize{Hairer-2-BGK} & 13.539 & 2.24 & 2.22 & 13.408 & 1.39 & 1.78 \\
			\scriptsize{John5} & 8.112 & 1.02 & 1.04 & 8.138 & 0.88 & 0.99 \\
			\scriptsize{Lazard-ascm2001} & 9.295 & 1.52 & 1.66 & 9.967 & 0.92 & 1.27 \\
			\scriptsize{Liu-Lorenz} & 25.120 & 1.4 & 1.46 & 25.239 & 1.25 & 1.37 \\
			\scriptsize{Mehta2} & 11.123 & 1.76 & 1.71 & 11.522 & 0.95 & 1.19 \\
			\scriptsize{Morgenstein} & 5.989 & 1.27 & 1.34 & 6.083 & 0.8 & 0.99 \\
			\scriptsize{Reif} & 74.261 & 2.41 & 2.43 & 346.306 & 3.05 & 3.18 \\
			\scriptsize{Xia} & 3.834 & 1.06 & 1.14 & 3.878 & 0.85 & 0.98 \\
			\scriptsize{Sys2161} & 6.714 & 2.01 & 2.09 & 26.213 & 3.35 & 3.49 \\
			\scriptsize{Sys2880} & 12.124 & 1.96 & 2.08 & 77.800 & 2.49 & 2.69 \\
			\scriptsize{Sys3054} & 13.568 & 2.25 & 2.28 & 13.510 & 1.38 & 1.77 \\
			\scriptsize{Sys3064} & 9.295 & 1.66 & 1.66 & 9.997 & 0.94 & 1.17 \\
			\scriptsize{Sys3068} & 25.723 & 1.13 & 1.45 & 25.331 & 1.28 & 1.08 \\
			\scriptsize{Sys3073} & 5.946 & 1.26 & 1.38 & 6.115 & 0.83 & 0.99 \\
			\scriptsize{Sys3170} & 3.836 & 1.06 & 1.12 & 3.898 & 0.82 & 0.99 \\
			\scriptsize{Sys3185} & 11.246 & 1.8 & 1.81 & 11.471 & 0.93 & 1.19 \\
			\scriptsize{Sys3238} & 13.531 & 2.23 & 2.2 & 13.459 & 1.36 & 1.77 \\
			\scriptsize{Sys3258} & 13.071 & 1.4 & 1.43 & 13.753 & 0.93 & 1.13 \\
			\scriptsize{Sys3306} & 9.316 & 1.67 & 1.7 & 9.964 & 0.95 & 1.17 \\
			\hline
		\end{tabular}
		\caption{Kalkbrener}\label{table:kalkexamples}
	\end{subtable}
	\caption{Some non-trivial examples solved in either the Lazard-Wu or Kalkbrener sense, with their serial runtime 
		and speed-up factors for different parallel configurations.}\label{table:nontrivialexamples}
\end{table*}

We test the various configurations of our implementation by considering
a suite of approximately 3000 real-world polynomial systems provided by MapleSoft.
Many of the examples lack opportunities for parallelism
as computations never split during their solving.
We have therefore selected
a subset of 340 systems which are either common
examples from the literature, or show potential for parallelism based on
computations splitting.
The experiments were run on a node with 2x6-core Intel
Xeon X5650 processors at 2.67GHz with
32KB L1 data ache, 256KB L2 cache, 12288KB L3 cache, and 48 GB of DDR3 RAM at 1333MHz.

We tested the selected systems on 12 different configurations: (Level vs.
Bubble algorithm) $\times$ (Lazard-Wu vs. Kalkbrener decomposition) $\times$ (serial (S),
coarse-grained (C) or coarse- and fine-grained (C+F) parallelism). 
We call one of serial, coarse-grained, or coarse- and fine-grained, the \textit{parallel configuration}.
See Section~\ref{sec:bubblevslevel} for details on the specifics of each type of parallelism.

Consider first the serial performance of the Level and 
Bubble algorithms for both Lazard-Wu (Table~\ref{table:lazexamples}) and Kalkbrener (Table~\ref{table:kalkexamples})
decomposition, to highlight their algorithmic differences.
Recall from Section~\ref{sec:Triangularize} that Level prevails over Bubble
in systems with many redundant components, while Kalkbrener decomposition can
be much faster than Lazard-Wu decomposition when deep levels of recursion are avoided.
%
These tables show that in some examples, such as John5, 
the Kalkbrener times can be dramatically faster thanks to avoiding expensive, deep recursive calls.
We can also see in systems Reif, Sys2161, and Sys2880, that the intermediate removal of redundant
components provided by the Level algorithm dramatically improves performance.
The systems 8-3-config-Li and Sys3258 show an interesting variant of this;
Lazard-Wu decomposition sees a significant benefit from removing intermediate redundant components (i.e. Level),
but this does not occur in the Kalkbrener decomposition where redundant components, and even whole branches,
can be avoided thanks to the height bound.
%

\begin{table}[h!tb]
	\renewcommand{\arraystretch}{1.15}
	\begin{tabular}{|c|c|c|c|}
		\hhline{~~|-|-|}
		\multicolumn{2}{r|}{}& {Bubble}&Level\\
		\hline
		\textit{Lazard-Wu} & S& 10 & 4 \\
		& C& 181 & 211 \\
		& C+F& 149 & 122\\
		\hline
		\textit{Kalkbrener} & S& 15 & 6 \\
		& C& 3 & 260 \\
		&C+F& 323 & 73\\
		\hline
	\end{tabular}
	\caption{For each configuration of Lazard-Wu/Kalkbrener and Bubble/Level, the number of systems which are solved fastest by each parallel configuration.}\label{table:bestconfigcounts}
\end{table}


\begin{figure*}[htb]
	\vspace{-1.02em}
	\centering
	\begin{subfigure}[t]{0.42\textwidth}
		\includegraphics[width=\textwidth]{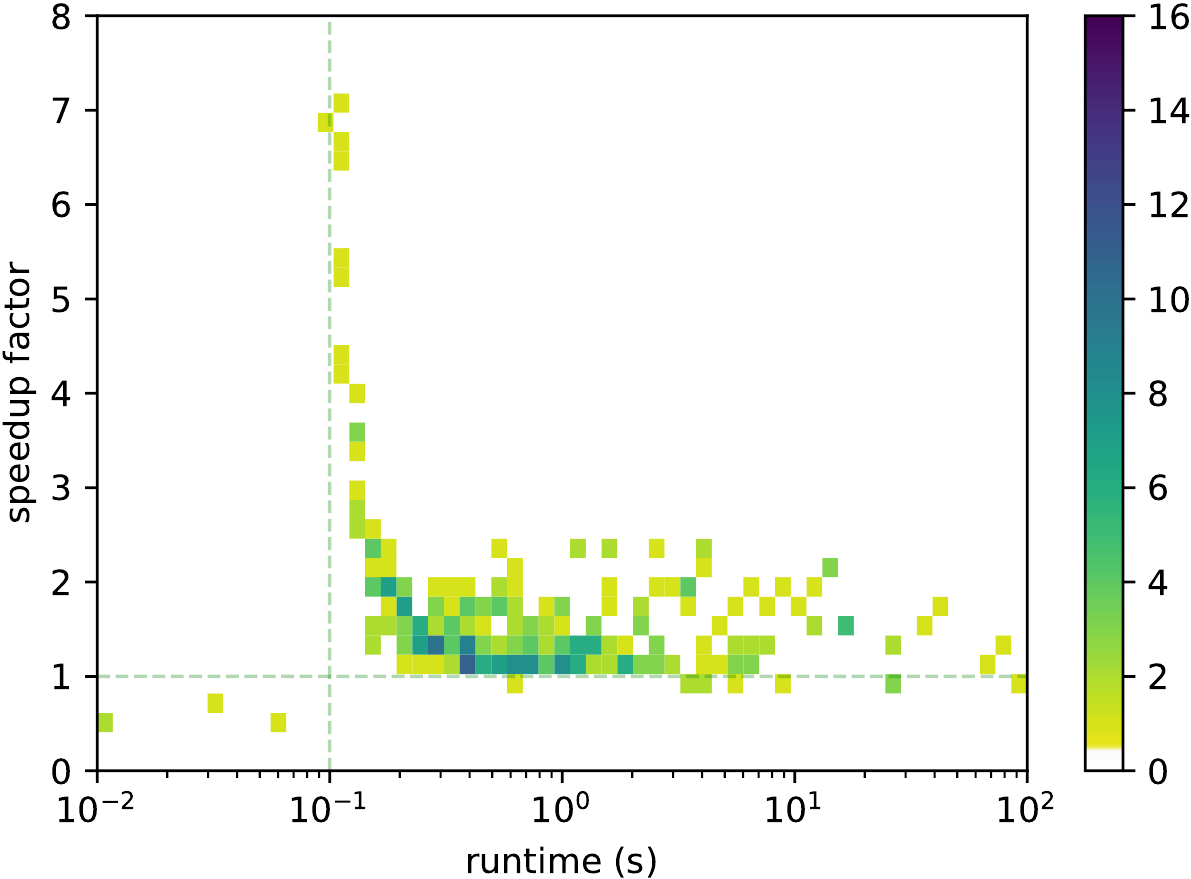}
		\caption{Lazard-Wu, \textsf{TriangularizeLevel}}\label{fig:histogram-LL}
	\end{subfigure}
	\hspace{4em}
	\begin{subfigure}[t]{0.42\textwidth}
		\includegraphics[width=\textwidth]{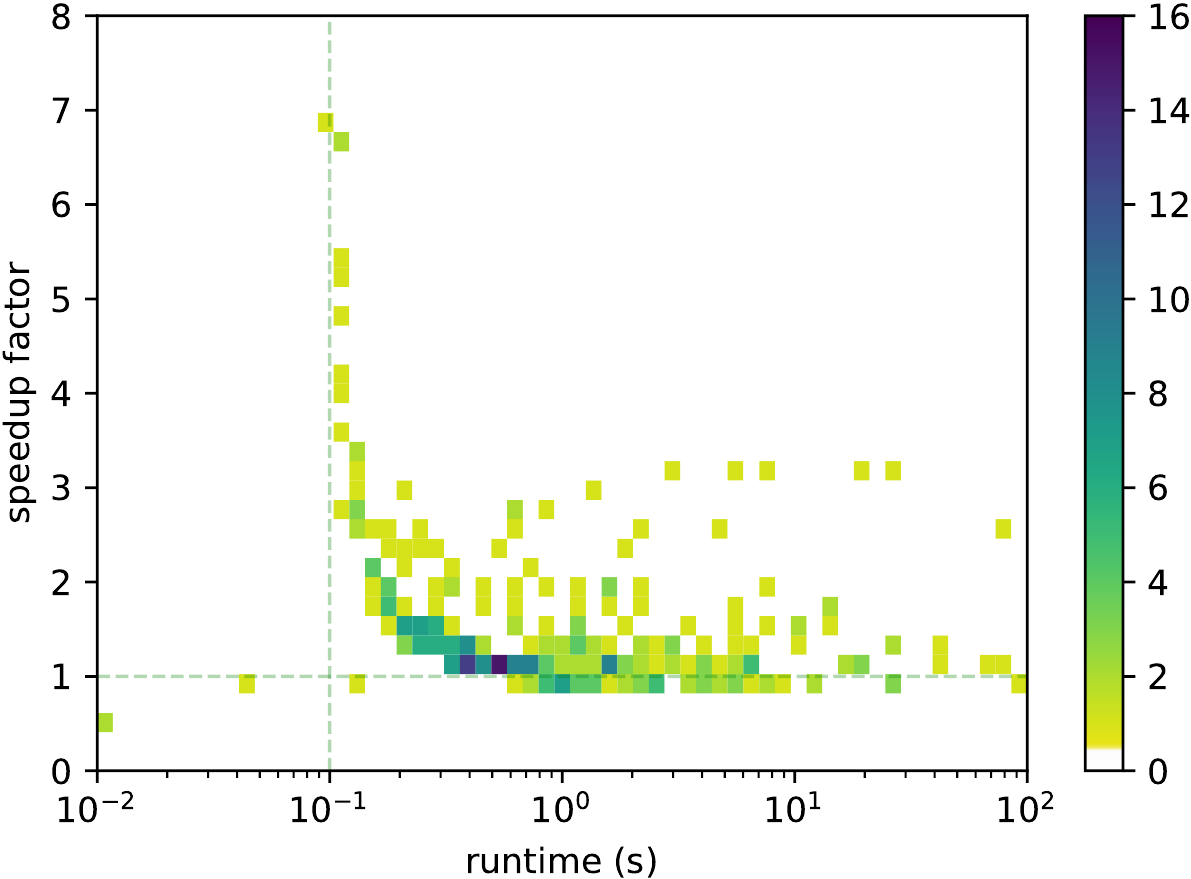}
		\caption{Lazard-Wu, \textsf{TriangularizeBubble}}\label{fig:histogram-BL}
	\end{subfigure} \\[1em]
	\begin{subfigure}[t]{0.42\textwidth}
		\includegraphics[width=\textwidth]{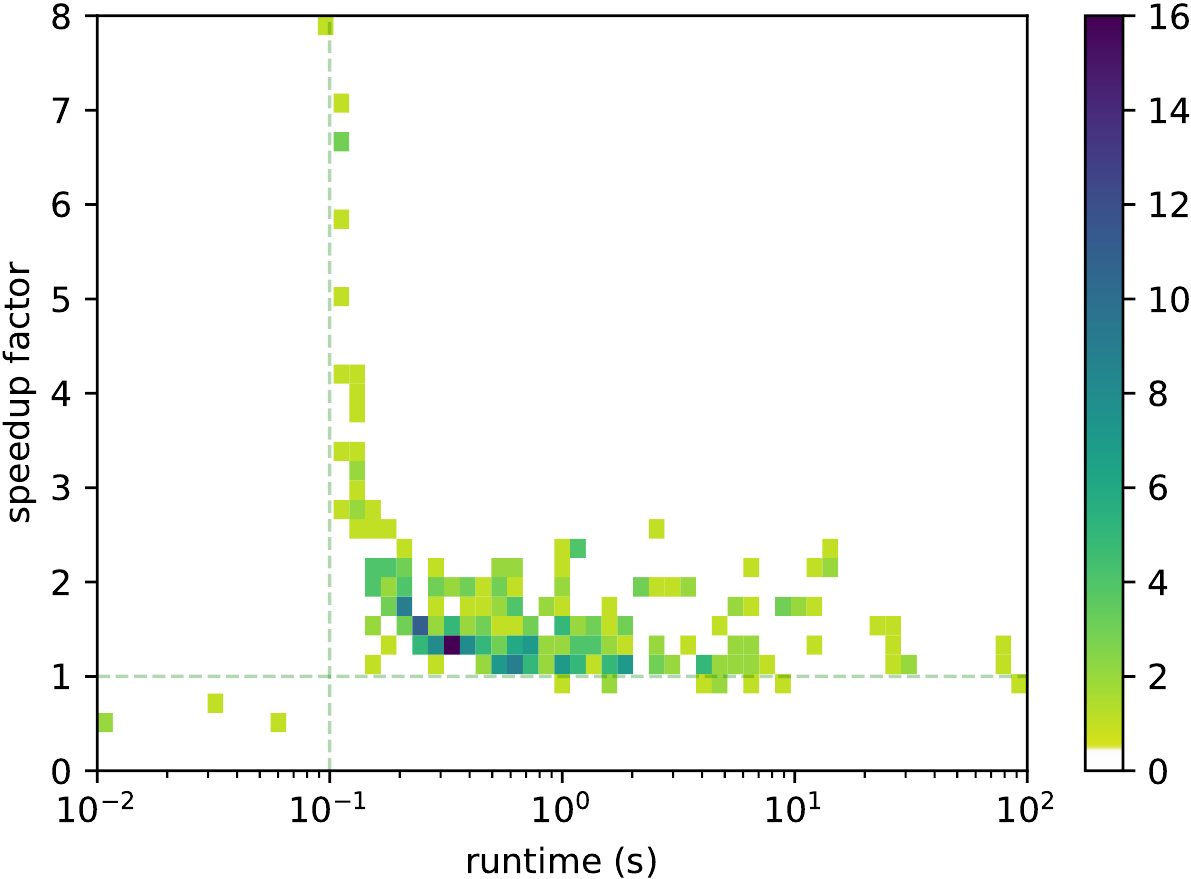}
		\caption{Kalkbrener, \textsf{TriangularizeLevel}}\label{fig:histogram-LK}
	\end{subfigure}
	\hspace{4em}
	\begin{subfigure}[t]{0.42\textwidth}
		\includegraphics[width=\textwidth]{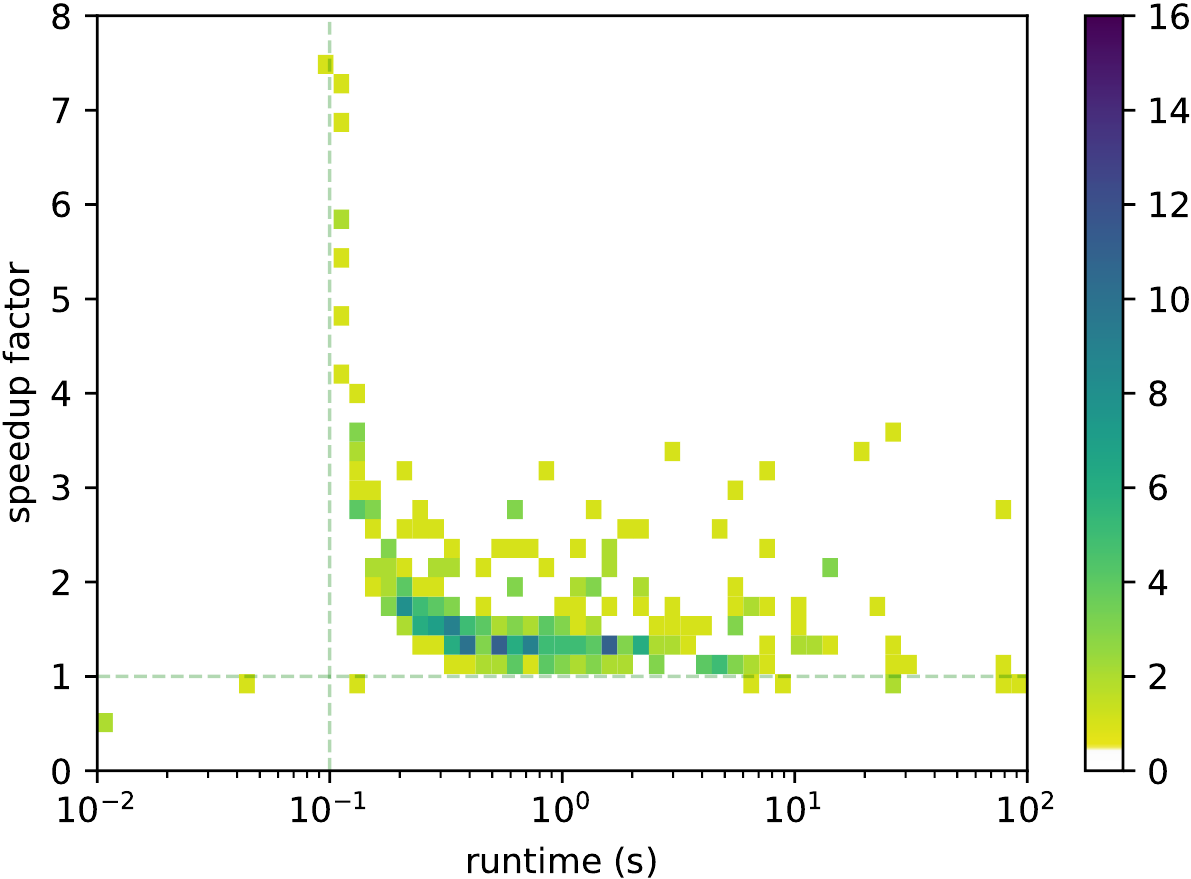}
		\caption{Kalkbrener, \textsf{TriangularizeBubble}}\label{fig:histogram-BK}
	\end{subfigure}
	\vspace{-0.5em}
	\caption{Histograms illustrating the distribution of runtime and speed-up factor of 340 examples solved by one choice of Lazard-Wu/Kalkbrender and one choice of \textsf{TriangularizeLevel}/\textsf{TriangularizeBubble}. These configurations show and use only coarse-grained parallelism---except (d) which also uses fine-grained---as this was the overall best configuration for speed-up. }\label{fig:histograms}
\end{figure*}

The remaining table, Table~\ref{table:bestconfigcounts}, and the
histogram plots in Figure~\ref{fig:histograms} illustrate patterns in
the 340 examples examined. Specifically,
Table~\ref{table:bestconfigcounts} shows a count of how many systems
were solved most quickly in a particular parallel configuration,
depending on the variant of the {\sf Triangularize} algorithm and
decomposition method used. The histograms show the number of systems
as a function of speed-up factor and runtime for a particular
variant of the {\sf Triangularize} algorithm, decomposition method and parallel
configuration (the best overall parallel configuration of the three is
shown).

Briefly, we see that \textsf{TriangularizeBubble}
admits more parallelism than \textsf{TriangularizeLevel}
and that solving in the Kalkbrener sense is more receptive
to our parallelization.
That is to say, they benefit more often from the
parallelism, in particular, from the addition of fine-grained.
Both of these can be seen from an upward shift of
the density in the plots and the counts presented in Table~\ref{table:bestconfigcounts} (Bubble to Level, left-to-right, and Lazard-Wu to Kalkbrener up-to-down).
We also see that \textsf{TriangularizeLevel} has fewer slow-downs overall.
We attribute this to the lack of overhead and contention
caused by the fine-grained parallelism, as we now explain.

\subsection{Parallel Performance: Bubble vs. Level}


The level-wise computation of \textsf{TriangularizeLevel}
presents challenges for parallelism as it creates several synchronization points
when it intermittently calls \textsf{RemoveRedundantComponents}.
This is an important optimization to the triangular decomposition algorithm
(as we have already seen from the systems
Reif, Sys2161, and Sys2880 in Table~\ref{table:nontrivialexamples})
but is detrimental to parallelism, particularly where 
the coarse-grained parallelism (i.e. the branches of computation)
is unbalanced at a particular level.
%
%
Table~\ref{table:bestconfigcounts} shows that
coarse-grained parallelism alone provides the best performance
a majority of the time for the Level algorithm.
By the multiple synchronizations required from the Level method, 
the benefit of components asynchronously bubbling up from
the subroutines via generators has no benefit to coarse-grained parallelism.
That being said, the use of generators may still help
subroutines work in parallel,
reducing the amount of time to compute, say, a single branch that dominates
the running time at a particular level. 
Effectively, it is helping to load-balance the coarse-grained tasks.
Where Kalkbrener decompositions can be naturally more balanced, due to 
their avoidance of deep recursive calls, we see from Table~\ref{table:bestconfigcounts}
fine-grained parallelism 
helps coarse-grained parallelism only 22\% of the time. 
Meanwhile, for Lazard-Wu decompositions, fine-grained
parallelism helps the computation 37\% of the time.

\textsf{TriangualarizeBubble}, in contrast, lacks synchronization points.
Further, the producer-consumer scheme between \textsf{Intersect} and \textsf{Triangularize}
(i.e. the coarse-grained parallelism) aligns directly 
with the producer-consumer scheme of the underlying subroutines (i.e. generators, the fine-grained
parallelism).
One could say that flow of data in the low-level
routines allows components to ``bubble up'' to \textsf{Triangularize}
earlier than they normally could,
promoting further coarse-grained parallelism.
Hence, the bubble method is generally more admissible to parallelization.
%
%

Considering all of this, there is no clear best configuration for solving problems in general.
The experimental data presented here bears the same information.
We can, however, observe some trends.
From Table~\ref{table:bestconfigcounts}
we see that, for Bubble, C+F is the clear winner for Kalkbrener decompositions.
Lazard-Wu decompositions see mixed results using this algorithm.
For Lazard-Wu decompositions it is more likely that computations will proceed
to deeper levels of recursion, which can, in some instances, consume all threads available
in the thread pool (see Section~\ref{sec:fineparallel})
and can cause other (coarse-grained) branches to effectively be serial in their
calls to subroutines.
Thus, in the cases where deep recursion does not occur, the computation can
indeed benefit from the fine-grained parallelism to 
start a different---and hopefully intensive---computation as early as possible.
In contrast, the tree-pruning nature of Kalkbrener decompositions make them 
less likely to reach deep levels
of recursion and thus see a better benefit from the subroutine generators.

\section{Conclusions and Future Work}
\label{sec:future}

Triangular decomposition of polynomial systems presents an interesting
challenge to parallel processing. The algorithm's ability to work
concurrently does not depend on the algorithm itself but rather on the 
problem instance.
Despite this we have presented two different parallel
algorithms for triangular decomposition which are adaptive to the
geometry of the problem currently being solved.

The Bubble algorithm admits the best parallelism, where coarse- and
fine-grained parallelism via asynchronous generators work together for
further task parallelism.  However, this method can be
hindered by redundant components.  The Level algorithm fixes the issue
with redundant components but is hindered by decreased parallelism
via multiple synchronization points.
In either case, fine-grained parallelism is only so effective.
Likely, data movement between threads is a limiting factor;
polynomials, and thus regular chains, can be very large objects,
particularly due to the \textit{expression swell} common in many
symbolic computations.

Despite these limitations we have performed parallel triangular decompositions
on hundreds of real-world systems and have obtained speed-up factors 
up to 8 on a 12-core machine. 
In the future we hope to build on this success. 
We hope to devise a scheme which intermittently 
removes redundant components but does so without requiring synchronization;
a mixture of both Level and Bubble.
This method would
likely follow a dynamic tree-pruning technique (beyond the pruning resulting
from Kalkbrener decompositions). We also hope to determine
better strategies for dynamically deciding between 
serial and asynchronous use of the low-level generators.

\subsection*{Acknowledgements}
The authors would like to thank IBM Canada Ltd (CAS project 880) and
NSERC of Canada (CRD grant CRDPJ500717-16).

%
%

\bibliographystyle{ACM-Reference-Format}
\bibliography{references}
\end{document}